\documentclass[11pt,a4paper]{amsart}

\usepackage{amsfonts,amssymb,amsmath,eucal,pinlabel,picins,array,hhline,picins}
\usepackage[all]{xy}
\usepackage{tabulary}
\usepackage{fancyhdr}

\newtheorem{theorem}{Theorem}[section]
\newtheorem*{Kast}{Kasteleyn's Theorem}
\newtheorem{lemma}[theorem]{Lemma}
\newtheorem{proposition}[theorem]{Proposition}
\newtheorem{corollary}[theorem]{Corollary}
\theoremstyle{definition}\newtheorem{definition}{Definition}
\theoremstyle{remark}\newtheorem*{example}{Example}
\theoremstyle{remark}\newtheorem*{remark}{Remark}

\newenvironment{romanlist}
        {\begin{enumerate}
        }
        {\end{enumerate}}

\newcounter{ticklistc}

\newcommand{\Z}{\mathbb Z}
\newcommand{\R}{\mathbb R}

\newcommand{\K}{\mathcal K}

\newcommand{\Q}{\mathcal Q}
\newcommand{\M}{\mathcal M}
\renewcommand{\S}{\mathcal S}

\newcommand{\e}{\varepsilon}

\newcommand{\G}{\Gamma}
\newcommand{\SI}{\Sigma}
\newcommand{\Pf}{\mathrm{Pf}}
\newcommand{\Arf}{\mathrm{Arf}}

\begin{document}

\title{A generalized Kac-Ward formula}

\author{David Cimasoni}   
\address{ETH Z\"urich, Departement Mathematik, R\"amistrasse 101, 8092 Z\"urich, Switzerland}
\email{david.cimasoni@math.ethz.ch}

\begin{abstract}
The Kac-Ward formula allows to compute the Ising partition function on a planar graph $G$ with straight edges from the determinant of a matrix of size $2N$,
where $N$ denotes the number of edges of $G$. In this paper, we extend this formula to any finite graph: the partition function can be written as an alternating
sum of the determinants of $2^{2g}$ matrices of size $2N$, where $g$ is the genus of an orientable surface in which $G$ embeds.
We give two proofs of this generalized formula. The first one is purely combinatorial, while the second relies on the Fisher-Kasteleyn reduction of the Ising model
to the dimer model, and on geometric techniques. As a consequence of this second proof, we also obtain the following fact:
the Kac-Ward and the Fisher-Kasteleyn methods to solve the Ising model are one and the same.
\end{abstract}

\maketitle

\pagestyle{myheadings}
\markboth{David Cimasoni}{A generalized Kac-Ward formula}


\section{Introduction}
\label{sec:intro}

The ``algebraic method" discovered in the 40's by Onsager and Kaufman~\cite{Ons,Kau} to solve the Ising model on a square lattice $G$ was widely considered to be
extremely difficult. This motivated Kac and Ward~\cite{K-W} (and later, Potts and Ward~\cite{P-W}) to try to find a more direct ``combinatorial method", building
on van der Waerden's reformulation of the Ising model~\cite{vdW}. They defined a matrix $M(G)$, with rows and columns indexed by oriented edges of $G$, whose determinant
coincides with the square of $Z^I(G)$, the Ising partition function on $G$. Unfortunately, many arguments in~\cite{K-W,P-W} are of heuristic nature, and some key
topological statement later turned out not to hold~\cite{She}.

Since the technicalities raised by a rigorous proof of the Kac-Ward formula seemed formidable, the focus shifted to finding combinatorial methods not involving directly the
Kac-Ward matrix $M(G)$. This was achieved independently and almost simultaneously by Sherman~\cite{She}, Hurst-Green~\cite{H-G}, Kasteleyn~\cite{Ka1} and Fisher~\cite{Fi1}.
Sherman built on (unpublished) ideas of Feynman to express $Z^I(G)$ as a formal infinite product of polynomials, for any planar graph $G$ with vertices of degree
2 or 4. The other three authors all essentially obtained the following result: they related $Z^I(G)$ with the dimer model partition function $Z^D(\G_G)$
on an associated graph $\G_G$, and then found a skew-symmetric adjacency matrix $A(\G_G)$ for $\G_G$ whose Pfaffian gives $Z^D(\G_G)$. This ``Pfaffian method" was later
extended by Kasteleyn~\cite{Ka2,Ka3} to any planar graph.

To the best of our knowledge, the first direct combinatorial proof of the Kac-Ward formula $Z^I(G)^2=\det(M(G))$ for any planar graph $G$
with straight edges was only obtained in 1999 by Dolbilin {\em et al.\/}~\cite{DZMSS}.

\medskip

In this paper, we present a generalized Kac-Ward formula, valid for any finite graph $G$ (possibly non planar, disconnected, with loops and multiple edges).
More precisely, let $\SI$ be an oriented closed surface of genus $g$ in which $G$ embeds. For each spin structure $\lambda\in\S(\SI)$, we define a generalized
Kac-Ward matrix $M^\lambda(G)$ whose rows and columns are indexed by oriented edges of $G$ (see Definition~\ref{def:KW}). Our formula reads
\[
Z^I(G)=\frac{1}{2^g}\sum_{\lambda\in \S(\SI)}(-1)^{\Arf(\lambda)}\det(M^\lambda(G))^{1/2},
\]
where $\Arf(\lambda)\in\Z_2$ is the Arf invariant of the spin structure $\lambda$ (Theorem~\ref{thm:KW}). If $G$ is a planar graph with straight edges, this formula
is precisely the classical Kac-Ward formula $Z^I(G)^2=\det(M(G))$. Using Bass's theorem on the Ihara-Selberg zeta function for graphs~\cite{Bas},
we immediately obtain a generalized Feynman-Sherman type formula expressing $Z^I(G)$ as an alternating sum of $2^{2g}$ formal infinite products (Corollary~\ref{cor:She}).
This slightly extends a result of Loebl~\cite{Loe} (from the case with all vertices of degree 2 or 4 to the general case),
and expresses it in what we believe to be the right language (spin structures instead of Sherman rotation numbers).

Our generalized Kac-Ward formula is new and original, but its mere existence is by no mean a surprise.
First, it is known since Dolbilin {\em et al.\/}~\cite{D96}, Tesler~\cite{Tes}, Gallucio-Loebl~\cite{G-L} and
Cimasoni-Reshetikhin~\cite{C-RI} that the dimer partition function $Z^D(\G)$ on any finite graph $\G$ can be expressed as the same alternating sum of $2^{2g}$ Pfaffians
of skew-symmetric adjacency matrices $A^\lambda(\G)$; by the correspondence $G\mapsto\G_G$, such a formula holds for $Z^I(G)$ as well.
Also, as mentioned above, an analogous Feynman-Sherman type formula was already obtained by Loebl~\cite{Loe}.
Finally, Masbaum informed the author that he independently obtained Corollary~\ref{cor:She}, the improved version of this Feynman-Sherman formula where spin structures replace
Sherman rotation numbers. He was therefore aware (via Bass's Theorem) of our formula.

However, we would like to emphasize the fact that we give two completely independent proofs of our main result.
The first one is of combinatorial nature, following the line of Dolbilin {\em et al.\/}~\cite{DZMSS}. (These authors developed most of the necessary tools to obtain the general
formula, but without the understanding of the importance of spin structures, one cannot define the right generalized Kac-Ward matrices.)
The second proof is of geometric nature. We first use the Fisher correspondence $G\mapsto\G_G$ together with the geometric treatment of the dimer model~\cite{C-RI}
to obtain the formula for $Z^I(G)$ in terms of the Pfaffians of $2^{2g}$ Kasteleyn matrices $A^\lambda(\G_G)$. (This was done in an analogous way in ~\cite{D96}.)
Then, we explicitly transform these Kasteleyn matrices into the corresponding generalized Kac-Ward matrices $M^\lambda(G)$,
using moves that do not change the determinant. Therefore, not only do we generalize the Kac-Ward approach and clarify the Feynman-Sherman one, but we also unify these two approaches with the Fisher-Kasteleyn Pfaffian method of~\cite{H-G,Ka1,Fi1,D96}. The fact that the Kac-Ward and Pfaffian methods are one and the same is by no mean obvious,
and they are very often considered as distinct (see e.g \cite{Zin,Loe}).

\medskip

The paper is organized as follows. In Section~\ref{sec:result}, we introduce the generalized Kac-Ward matrices (Definition~\ref{def:KW}), state our main result
(Theorem~\ref{thm:KW}) and its consequence (Corollary~\ref{cor:She}). In Section~\ref{sec:comb}, we give the combinatorial proof, following the line of~\cite{DZMSS},
and assuming many of their results. (In this sense, the combinatorial proof given here is not entirely self-contained.) Finally, Section~\ref{sec:geo} deals with
the geometric proof. We begin by giving an improved version of the Fisher construction $G\mapsto\G_G$ (Subsection~\ref{sub:Fis}). We then recall the geometric approach
to the dimer model (Subsection~\ref{sub:dimers}) and finally show how to identify the Kac-Ward and Fisher-Kasteleyn matrices (Subsection~\ref{sub:KWvsKF}).

\subsection*{Acknowledgments}
The author would like to express his thanks to Martin Loebl and Gregor Masbaum for stimulating discussions.


\section{Statement of the formula}
\label{sec:result}

Let $G$ be a finite graph, possibly disconnected, possibly with loops and multiple edges. Let us associate to each edge $e\in E(G)$ a formal variable $x_e$.
As discovered by van der Waerden \cite{vdW}, the partition function for the Ising model on $G$ with zero magnetic field can be put in the form
\[
Z^I(G)=\sum_{G'\subset G}\prod_{e\in E(G')}x_e,
\]
where the sum is over all subgraphs $G'$ of $G$ such that each vertex of $G$ is met by an even number of edges of $G'$. In other words, the sum is over the set
\[
Z_1(G;\Z_2)=\big\{\xi=\textstyle{\sum_e\xi_e\,e\in C_1(G;\Z_2)}\,|\,\partial\xi=0\in C_0(G;\Z_2)\big\}
\]
of 1-cycles modulo 2 in $G$. Therefore, if we denote by $|\xi|$ the set of edges of $G$ such that $\xi_e$ is equal to 1, the Ising partition function on $G$ can be written
\[
Z^I(G)=\sum_{\xi\in Z_1(G;\Z_2)}x(|\xi|), \; \text{ where } \; x(|\xi|)=\prod_{e\in|\xi|}x_e. 
\]

The aim of this section is to state a closed formula for this partition function in terms of the determinant of $2^{2g}$ ``generalized Kac-Ward matrices'',
where $g$ denotes the genus of a surface where the graph $G$ embeds. These matrices will be defined using spin structures.
We shall therefore start by recalling the main features of these geometric objects.

Let $\SI$ be a closed oriented smooth surface endowed with a Riemannian metric, and let $F\SI\to\SI$ denote its orthogonal frame bundle.
A {\em spin structure\/} on $\SI$ is a cohomology class $\lambda\in H^1(F\SI;\Z_2)$ whose restriction to each fiber $F_x$ of this principal $S^1$-bundle is the
non-trivial element of $H^1(F_x;\Z_2)=\Z_2$. In other words, a spin structure assigns to each framed curve in $\SI$ an integer modulo 2, in such a way that
the following two conditions are satisfied: the boundary of any disc with constant framing has value 0, and the boundary of any disc with framing induced by an outgoing
vector field has value 1. In particular, a spin structure allows to compute a modulo 2 winding number for any closed curve $\gamma$ in $\SI$: simply consider the value assigned
to $\gamma$ with framing induced by the tangent vector field along $\gamma$. We shall denote this number by $\lambda(\vec{\gamma})\in\Z_2$.

One easily checks that the set $\S(\SI)$ of spin structures on $\SI$ is an affine $H^1(\SI;\Z_2)$-space.
In particular, there are exactly $2^{2g}$ distinct spin structures on a closed oriented surface $\SI$ of genus $g$.
A convenient way to give oneself a spin structure on $\SI$ is via a vector field with isolated zeroes of even index: the associated spin structure will be
obtained by computing the modulo 2 winding number of this vector field along a closed framed curve, with respect to the framing of this curve. The evenness of the indices
ensures that both conditions described above are satisfied.

Recall finally that spin structures on a surface $\SI$ come in two classes, determined by their ``Arf invariant": the even spin structures, with Arf invariant 0,
and the odd ones, with Arf invariant 1 (see the definition in Subsection~\ref{sub:Joh}). These two classes coincide with the orbits of
the action of the diffeomorphism group of $\SI$ on $\S(\SI)$.

We are now ready to define the generalized Kac-Ward matrices.

Let $G$ be a finite graph, and let $E$ denote the set of {\em oriented\/} edges of $G$. For any $e\in E$, we shall denote by $s(e)$ its starting point, by $f(e)$ its endpoint,
and by $-e$ the same edge with the opposite orientation. By abuse of notation, we shall denote by $x_e=x_{-e}$ the formal variable associated to the unoriented edge
corresponding to $e$ and $-e$. Now, embed $G$ in a closed oriented surface $\SI$, and fix a spin structure $\lambda\in \S(\SI)$ via a vector field on $\SI$ with isolated
zeroes of even index in $\SI\setminus G$. Finally, fix an arbitrary point in each edge of $G\subset\SI$.

\begin{definition}
\label{def:KW}
The {\em generalized Kac-Ward matrix\/} associated to $G$ and $\lambda$ is the $|E|\times|E|$ matrix $I-T^\lambda(G)$,
where $I$ is the identity and $T^\lambda(G)$ denotes the transition matrix defined by
\[
T^\lambda(G)_{e,e'}=
\begin{cases}
\exp\left(\frac{i}{2}\alpha_\lambda(e,e')\right)\,x_e& \text{if $f(e)=s(e')$ but $e'\neq -e$;} \\
0 & \text{otherwise.}
\end{cases}
\]
Here, $\alpha_\lambda(e,e')\in\R$ is the rotation number of the tangent vector field along $e$ followed by $e'$ with respect to the vector field $\lambda$,
from the fixed point in the edge $e$ to the fixed point in the edge $e'$.
\end{definition}

Obviously, the matrix $I-T^\lambda(G)$ will depend on an ordering of the set $E$, on the choice of the vector field representing the spin structure $\lambda$, and
on the choice of the fixed points in the edges of $G$. However, its determinant will not. This is the first part of our main result. 

\begin{theorem}
\label{thm:KW}
Let $G$ be a finite graph embedded in an orientable closed surface $\SI$ of genus $g$. For any spin structure $\lambda\in \S(\SI)$, the determinant
of the matrix $I-T^\lambda(G)$ is the square of a polynomial in the variables $\{x_e\}_{e\in E(G)}$ which only depends on $\lambda$ and on $G$.
If $\det(I-T^\lambda(G))^{1/2}$ denotes the square root with constant coefficient equal to $+1$, then the Ising partition function on $G$ is equal to
\[
Z^I(G)=\frac{1}{2^g}\sum_{\lambda\in \S(\SI)}(-1)^{\Arf(\lambda)}\det(I-T^\lambda(G))^{1/2},
\]
where $\Arf(\lambda)\in\Z_2$ is the Arf invariant of the spin structure $\lambda$.
\end{theorem}

A remarkable feature of this formula is that the left-hand side only depends on the abstract graph $G$, while the right-hand side a priori depends on
the way this graph is embedded in a surface.

\begin{example}
Let $G$ be a planar graph with straight edges. Obviously, there is a single spin structure $\lambda$ in the plane, whose most natural representative is a constant
vector field. The associated rotation number $\alpha_\lambda(e,e')$ is simply given by the oriented angle $\sphericalangle(e,e')$ between the oriented edges $e$
and $e'$. Therefore, we have the formula
\[
Z^I(G)=\det(M(G))^{1/2},
\]
where $M(G)$ is the $|E|\times |E|$ matrix with coefficients
\[
M(G)_{e,e'}=\begin{cases}
1& \text{if $e=e'$;} \\ 
-\exp\left(\frac{i}{2}\sphericalangle(e,e')\right)\,x_e& \text{if $f(e)=s(e')$ but $e'\neq -e$;} \\
0 & \text{otherwise.}
\end{cases}
\]
This is precisely the Kac-Ward formula \cite{K-W}.
\end{example}

\begin{example}
Let $G$ be the ``figure-eight graph" given by a single vertex and two loops $e_1,e_2$ at this vertex with associated variables $x_1,x_2$.
Obviously, the Ising partition function is given by $Z^I(G)=1+x_1+x_2+x_1x_2$. We will now check that the formula in Theorem~\ref{thm:KW} gives the right result.

\parpic[r]{$\begin{array}{c}
\labellist\small\hair 2.5pt
\pinlabel {$e_2$} at 100 90
\pinlabel {$e_1$} at 260 90
\endlabellist
\includegraphics[height=1.2cm]{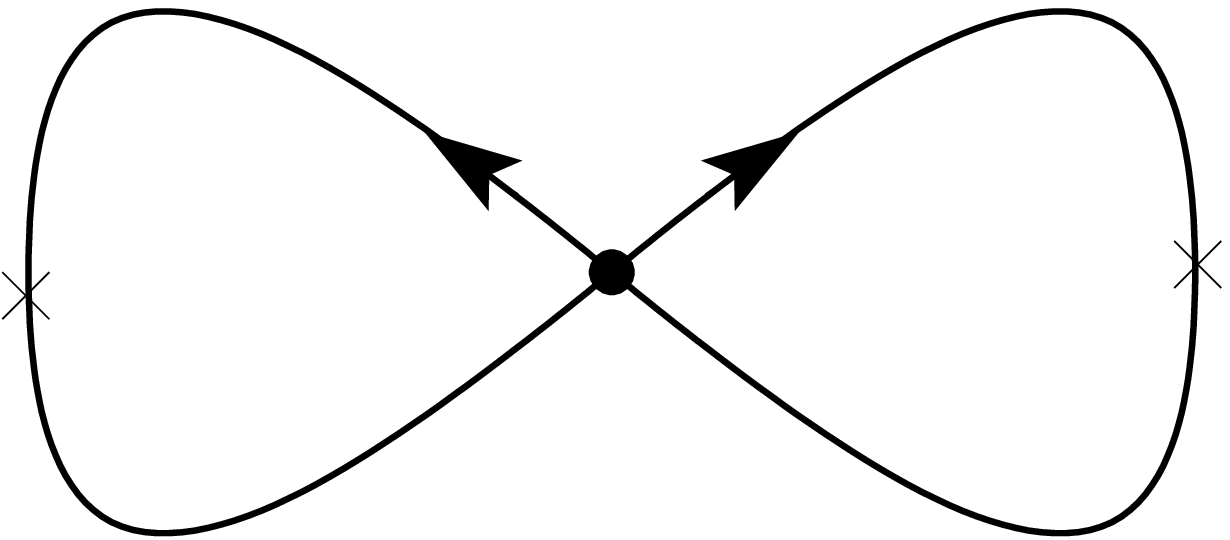}
\end{array}$}
Let us first consider $G$ embedded in the plane as illustrated opposite, with the marked fixed points in its edges. Let $\lambda$ be the unique spin structure on the plane,
given by a constant vector field. The corresponding rotation numbers are given by
\[
\alpha_\lambda(e_2,e_2)=-\alpha_\lambda(e_1,e_1)=2\pi,\quad\alpha_\lambda(e_1,e_2)=\alpha_\lambda(e_2,e_1)=0,\quad\text{\it etc...}
\]
If the set $E$ is ordered by $\{e_1,e_2,-e_1,-e_2\}$, then the corresponding Kac-Ward matrix is given by
\[
I-T^\lambda(G)=\begin{pmatrix}
1+x_1&-x_1&0&-ix_1\\
-x_2&1+x_2&ix_2&0\\
0&ix_1&1+x_1&-x_1\\
-ix_2&0&-x_2&1+x_2
\end{pmatrix}.
\]
Its determinant being equal to $(1+x_1)^2(1+x_2)^2$, we get the claimed equality $\det(I-T^\lambda(G))^{1/2}=Z^I(G)$.

\parpic[r]{$\begin{array}{c}
\labellist\small\hair 2.5pt
\pinlabel {$e_1$} at 40 80
\pinlabel {$e_2$} at 80 45
\pinlabel {$\gamma_1$} at 157 33
\pinlabel {$\gamma_2$} at 25 163
\endlabellist
\includegraphics[height=2.5cm]{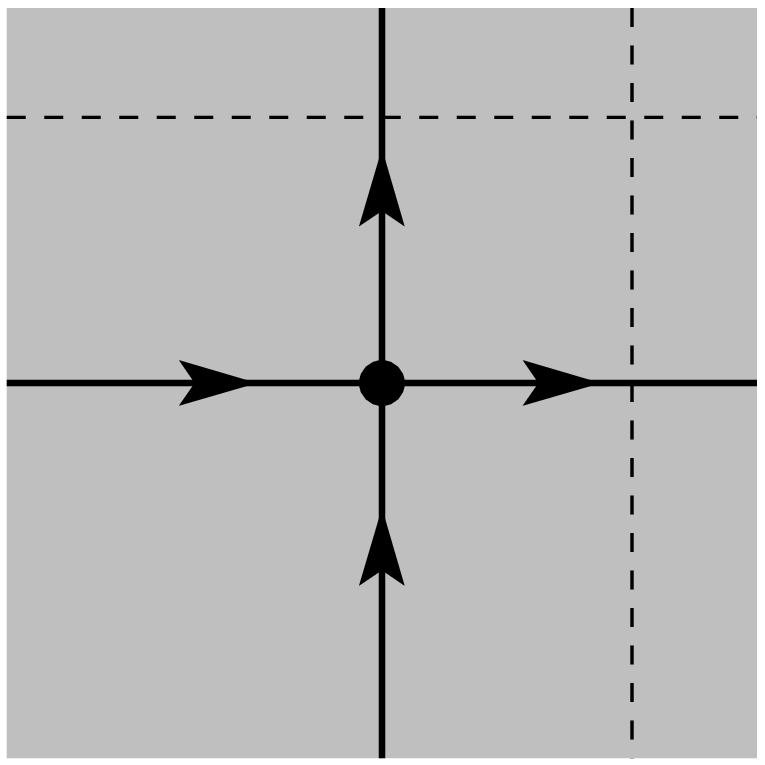}
\end{array}$}Let us now consider $G$ embedded in the torus $\mathbb{T}^2$ as illustrated opposite. Let $\lambda_{1,1}$ denote the vector field on $\mathbb{T}^2$
induced by a constant vector field on the plane, and let $\lambda_{1,-1}$ (resp. $\lambda_{-1,1}$, $\lambda_{-1,-1}$) be obtained from $\lambda_{1,1}$ via a full twist
along the closed curve $\gamma_1$ (resp. $\gamma_2$, $\gamma_1$ and $\gamma_2$). Using the notation $\omega:=\exp(i\pi/4)$, we obtain
\begin{align*}
\det(I-T^{\lambda_{\e_1,\e_2}}(G))&=\begin{vmatrix}
1-\e_1x_1&-\e_2\omega x_1&0&-\overline{\omega}x_1\\
-\e_1\overline{\omega}x_2&1-\e_2x_2&-\omega x_2&0\\
0&-\e_1\e_2\overline{\omega}x_1&1-\e_1x_1&-\e_1\omega x_1\\
-\e_1\e_2\omega x_2&0&-\e_2\overline{\omega}x_2&1-\e_2x_2
\end{vmatrix}\\
&=(1-\e_1x_1-\e_2x_2-\e_1\e_2x_1x_2)^2.
\end{align*}
Using the definition of the Arf invariant (see Subsection~\ref{sub:Joh}), one gets
\[
(-1)^{\Arf(\lambda_{\e_1,\e_2})}=\frac{1}{2}(1-\e_1-\e_2-\e_1\e_2)=\begin{cases}-1 & \text{if $\e_1=\e_2=1$;} \\
\phantom{-}1 & \text{otherwise.}\end{cases}
\]
Hence, the alternating sum in Theorem~\ref{thm:KW} is again equal to
$Z^I(G)$ as claimed.
\end{example}

We now turn to a consequence of Theorem~\ref{thm:KW}.

Let $\gamma$ denote an oriented closed path on a graph $G$. We will say that $\gamma$ is {\em reduced} if it never backtracks, that is, if no oriented edge
$e$ is immediately followed by the oriented edge $-e$. The oriented closed path $\gamma$ will be called {\em prime} if, when viewed as a cyclic word,
it cannot be expressed as the product $\delta^r$ of a given closed path $\delta$ for any $r\ge 2$. Note that the oriented closed path $\gamma$ is reduced (resp. prime)
if and only if $-\gamma$ is.
It therefore makes sense to talk about prime reduced unoriented closed paths.
Finally, recall that a spin structure $\lambda\in\S(\SI)$ allows to compute a modulo 2 winding number for an oriented closed curve $\gamma$ in $\SI$,
which we denote by $\lambda(\vec{\gamma})\in\Z_2$. Note also that $\gamma$ and $-\gamma$ have the same winding numbers.

Applying Bass's Theorem~\cite{Bas,F-Z} to our generalized Kac-Ward matrices, we immediately obtain the following corollary, which extends results of Sherman~\cite{She}
and Loebl~\cite{Loe}.

\begin{corollary}
\label{cor:She}
Let $G$ be a finite graph embedded in an orientable closed surface $\SI$ of genus $g$. Then the Ising partition function on $G$ is equal to
\[
Z^I(G)=\frac{1}{2^g}\sum_{\lambda\in \S(\SI)}(-1)^{\Arf(\lambda)}\prod_{\gamma\in\mathcal{R}}\big(1-(-1)^{\lambda(\vec{\gamma})}x(\gamma)\big),
\]
where $\mathcal{R}$ denotes the (infinite) set of prime reduced unoriented closed paths on the graph $G$, and $x(\gamma)$ stands for $\prod_{e\in\gamma}x_e$. \qed
\end{corollary}

Gregor Masbaum informed the author that he independently obtained this corollary working directly with infinite products as in~\cite{She,Loe}.
Via Bass's Theorem, he therefore has yet another proof of Theorem~\ref{thm:KW}.


\section{The combinatorial proof}
\label{sec:comb}

The aim of this section is to give a first proof of Theorem~\ref{thm:KW}, of combinatorial nature. More precisely, we first use Johnson's theorem on spin structures
and some standard tricks to reduce our formula to a combinatorial statement (Proposition~\ref{prop:comb}).
We then build on some results of \cite{DZMSS} to prove this combinatorial statement.

\subsection{Computing the determinants using Johnson's theorem}
\label{sub:Joh}

We shall start by giving a proof of the first part of Theorem~\ref{thm:KW}, that is, that the determinant of $I-T^\lambda(G)$ only depends on $\lambda$ and $G$.
As we will see, it follows quite easily from a result of Johnson \cite{Joh}, that we now recall.

As above, let $\SI$ be a closed oriented surface of genus $g$. A {\em quadratic form\/} on $H_1(\SI;\Z_2)$ is a map $q\colon H_1(\SI;\Z_2)\to\Z_2$ such that
$q(x+y)=q(x)+q(y)+x\cdot y$ for all $x,y\in H_1(\SI;\Z_2)$, where $\cdot$ denotes the intersection form. The set $\Q(\SI)$ of such quadratic forms is clearly
an affine $H^1(\SI;\Z_2)$-space.
Given a vector field $\lambda$ representing a spin structure, let $q_\lambda\colon H_1(\SI;\Z_2)\to\Z_2$ denote the map defined as follows.
Represent a class $\alpha\in H_1(\SI;\Z_2)$ by a collection of disjoint oriented regular simple closed curves
$C_1,\dots,C_m$ in $\SI$ avoiding the zeroes of the vector field $\lambda$, and set
\[
q_\lambda(\alpha)= \sum_{i=1}^m (\lambda(\vec{C_i})+1) \pmod{2}
\]
where $\lambda(\vec{C_i})$ denotes the winding number of the tangential vector field along $C_i$ with respect to the vector field $\lambda$. Johnson's theorem asserts that this assignment $\lambda\mapsto q_\lambda$ defines an $H^1(\SI;\Z_2)$-equivariant bijection between the set $\S(\SI)$ of spin structures on $\SI$ and $\Q(\SI)$.
The {\em Arf invariant\/} of a spin structure $\lambda\in\S(\SI)$ is then defined as the Arf invariant of the corresponding quadratic form $q_\lambda$, that is, the modulo 2
integer $\Arf(\lambda)\in\Z_2$ given by the equality
\[
(-1)^{\Arf(\lambda)}=\frac{1}{2^g}\sum_{\alpha\in H_1(\SI;\Z_2)}(-1)^{q_\lambda(\alpha)}.
\]
We shall need a single property of this invariant, namely that it satisfies the equality
\begin{equation}
\label{equ:Arf}
\frac{1}{2^g}\sum_{\lambda\in\S(\SI)}(-1)^{\Arf(\lambda)}(-1)^{q_\lambda(\alpha)}=1
\end{equation}
for any $\alpha\in H_1(\SI;\Z_2)$. We refer to \cite[Lemma 2.10]{L-M} for the easy proof.

\medskip

Let us start the computation of the determinant
\[
\det(I-T^\lambda(G))=:\det M^\lambda=\sum_{\sigma\in S(E)}(-1)^{\text{sgn}(\sigma)}\prod_{e\in E}M^\lambda_{e,\sigma(e)}.
\]
Each permutation $\sigma\in S(E)$ decomposes into disjoint cycles, inducing a partition $E=\bigsqcup_j E_j(\sigma)$ of the set of oriented
edges into orbits of length $\ell_j(\sigma)$. The corresponding contribution in the determinant is
\[
(-1)^{\text{sgn}(\sigma)}\prod_{e\in E}M^\lambda_{e,\sigma(e)}=\prod_j(-1)^{\ell_j(\sigma)+1}\prod_{e\in E_j(\sigma)}M^\lambda_{e,\sigma(e)}.
\]
Note that $M^\lambda_{e,e}$ is equal to $1$ if $e$ is not a loop. Therefore, the non-loop oriented edges that are fixed by $\sigma$ do not contribute to this product,
and the corresponding orbits can be removed. By definition of $M^\lambda$, a permutation $\sigma\in S(E)$ will have a non-zero contribution if and only if each of
the remaining orbits $E_j(\sigma)$ forms an oriented closed path $\gamma_j(\sigma)$ in $G$, such that the corresponding set of closed paths
$\gamma(\sigma)=\{\gamma_j(\sigma)\}_j$ covers all oriented loops of $G$, and such that $\gamma(\sigma)$ is a so-called {\em admissible path\/}, that is:
\begin{romanlist}
\item{it passes through each edge of $G$ at most twice, and if so, in opposite directions;}
\item{it never backtracks, that is, no $\gamma_j(\sigma)$ contains two successive oriented edges of the form $e$ and $-e$.}
\end{romanlist}
The corresponding non-zero contributions in $\det M^\lambda$ will be of the form
\[
\prod_j(-1)^{\ell_j(\sigma)+1}\prod_{e\in E_j(\sigma)}\Big(\delta_{\ell_j(\sigma),1}-\exp\big(\textstyle{\frac{i}{2}}\alpha_\lambda(e,\sigma(e))\big)x_e\Big),
\]
where $\delta_{\ell_j(\sigma),1}$ is the Kronecker symbol. This is equal to
\[
\prod_j\Big(\delta_{\ell_j(\sigma),1}-\prod_{e\in E_j(\sigma)}\exp\left(\textstyle{\frac{i}{2}}\alpha_\lambda(e,\sigma(e))\right)x_e\Big).
\]
Let $\Gamma(G)$ denote the set of admissible closed paths $\gamma=\{\gamma_j\}_j$ in $G$, and $\widetilde\Gamma(G)$ the set of admissible closed paths that cover
all oriented loops of $G$. We have:
\begin{align*}
\det M^\lambda&=
\sum_{\gamma\in\widetilde\Gamma(G)}\prod_j\Big(\delta_{\ell_j,1}-\prod_{e\in\gamma_j}\exp\big(\textstyle{\frac{i}{2}}\alpha_\lambda(e,\sigma(e))\big)x_e\Big)\\
&=\sum_{\gamma\in\widetilde\Gamma(G)}\prod_j\Big(\delta_{\ell_j,1}-\exp\big(\textstyle{\frac{i}{2}}\alpha_\lambda(\vec{\gamma}_j)\big)x(\gamma_j)\Big),
\end{align*}
where $x(\gamma_j)=\prod_{e\in\gamma_j}x_e$ and $\alpha_\lambda(\vec{\gamma}_j)$ is the rotation number of the tangent vector field along $\gamma_j$ with respect to the
vector field $\lambda$. Using the notation $z(\gamma_j)=-\exp(\frac{i}{2}\alpha_\lambda(\vec{\gamma}_j))$, we get:
\begin{align*}
\det M^\lambda&=
\sum_{\gamma\in\widetilde\Gamma(G)}\prod_j(\delta_{\ell_j,1}+z(\gamma_j)x(\gamma_j))\\
&= \sum_{\gamma\in\widetilde\Gamma(G)}\prod_{\{j|\ell_j=1\}}(1+z(\gamma_j)x(\gamma_j))\prod_{\{j|\ell_j\neq 1\}}z(\gamma_j)x(\gamma_j)\\
&= \sum_{\gamma\in\widetilde\Gamma(G)}\Big(\sum_{S\subset\{k|\ell_k=1\}}\prod_{j\in S}z(\gamma_j)x(\gamma_j)\Big)\prod_{\{j|\ell_j\neq 1\}}z(\gamma_j)x(\gamma_j)\\
&= \sum_{\gamma\in\Gamma(G)}\prod_jz(\gamma_j)x(\gamma_j)\\
&=:\sum_{\gamma\in\Gamma(G)}z(\gamma)x(\gamma).
\end{align*}

We now claim that for any given admissible closed path $\gamma=\{\gamma_1,\dots,\gamma_m\}$ in $G$, we have the equality
\[
z(\gamma):=\prod_{j=1}^m -\exp(\textstyle{\frac{i}{2}}\alpha_\lambda(\vec{\gamma}_j))=(-1)^{q_\lambda(\gamma)+t(\gamma)},
\]
where $q_\lambda$ is the quadratic form associated to the spin structure $\lambda$, and $t(\gamma)$ denotes the number of double points of the path $\gamma$ perturbed in
general position. (Note that this number is well-defined, as $\gamma$ never backtracks.) Both sides of this equality are unchanged when the path is perturbed, so it is enough
to prove the corresponding equality for a family $\alpha=\{\alpha_1,\dots,\alpha_m\}$ of oriented closed curves in general position in $\SI$.
\parpic[r]{$\begin{array}{c}
\labellist\small\hair 2.5pt
\pinlabel {$\alpha$} at 20 20
\pinlabel {$\alpha'$} at 210 20
\endlabellist
\includegraphics[height=1.5cm]{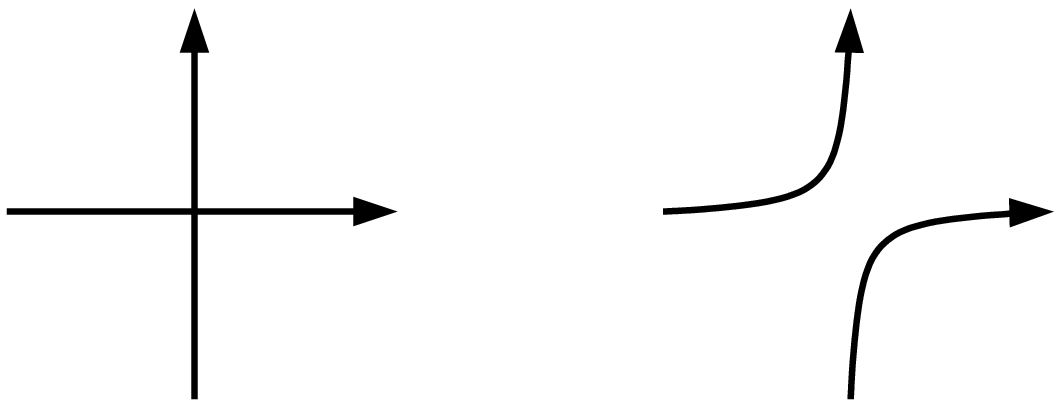}
\end{array}$}
This can be done by induction on the number $t(\alpha)$ of double points of $\alpha$. If $t(\alpha)$ is zero, $\alpha$ consists of a disjoint union of oriented simple closed
curves. In this case, the equality above is precisely Johnson's definition of the quadratic form $q_\lambda$. On the other hand, smoothing out one double point of
$\alpha$ as illustrated above transforms it into a new family $\alpha'$ of $m'=m\pm 1$ oriented closed curves with
$q_\lambda(\alpha')=q_\lambda(\alpha)$ and $t(\alpha')=t(\alpha)-1$. One easily checks that the left hand side of the equation displayed above will also be multiplied
by $-1$, thus proving the claim.

Summing up, we have proved the following: For any spin structure $\lambda\in\S(\SI)$, the determinant of the corresponding generalized Kac-Ward matrix is equal to
\[
\det(I-T^\lambda(G))=\sum_{\gamma\in\Gamma(G)}(-1)^{q_\lambda(\gamma)+t(\gamma)}x(\gamma).
\]
In particular, this determinant only depends on $\lambda\in\S(\SI)$ and on $G$.

\medskip

Given $X,Y$ two subgraphs of $G$ with no edge in common, let $\G(X,Y)$ denote the set of admissible paths that pass once on the edges of $X$, twice on the edges of
$Y$, and that do not pass on the edges of $G\setminus(X\cup Y)$. Obviously, this set is empty unless $X=|\xi|$ for some $\xi\in Z_1(G;\Z_2)$. Therefore:
\begin{align*}
\det(I-T^\lambda(G))&=\sum_{\gamma\in\Gamma(G)}(-1)^{q_\lambda(\gamma)+t(\gamma)}x(\gamma)\\
&=\sum_{\xi\in Z_1(G;\Z_2)}\sum_{Y}\sum_{\gamma\in\G(|\xi|,Y)}(-1)^{q_\lambda(\gamma)+t(\gamma)}x(|\xi|)x(Y)^2\\
&=\sum_{\xi\in Z_1(G;\Z_2)}(-1)^{q_\lambda(\xi)}\sum_{Y}\Big(\sum_{\gamma\in\G(|\xi|,Y)}(-1)^{t(\gamma)}\Big)x(|\xi|)x(Y)^2,
\end{align*}
where the second sum is over all subgraphs $Y$ of $G$ with no edge in common with $|\xi|$. On the other hand, let $Z_\lambda(G)$
denote the following twisted partition function
\[
Z_\lambda(G):=\sum_{\xi\in Z_1(G;\Z_2)}(-1)^{q_\lambda(\xi)}x(|\xi|).
\]
Its square is equal to
\begin{align*}
Z_\lambda(G)^2&= \sum_{\xi_1,\xi_2}(-1)^{q_\lambda(\xi_1)+q_\lambda(\xi_2)}x(|\xi_1|)x(|\xi_2|)\\
&= \sum_{\xi_1,\xi_2}(-1)^{q_\lambda(\xi_1+\xi_2)}(-1)^{\xi_1\cdot\xi_2}x(|\xi_1|)x(|\xi_2|)\\
&= \sum_{\xi\in Z_1(G;\Z_2)}(-1)^{q_\lambda(\xi)}\sum_{Y}\Big(\sum_{\genfrac{}{}{0pt}{}{\xi_1+\xi_2=\xi}{|\xi_1|\cap|\xi_2|=Y}}(-1)^{\xi_1\cdot\xi_2}\Big)x(|\xi|)x(Y)^2,
\end{align*}
where the second sum is over all subgraphs $Y$ of $G$ with no edge in common with $|\xi|$. Therefore, the combinatorial proof of Theorem~\ref{thm:KW} boils down to the following:

\begin{proposition}
\label{prop:comb}
For any $\xi\in Z_1(G;\Z_2)$ and any subgraph $Y$ of $G$ with no edge in common with $|\xi|$, we have the equality
\[
\sum_{\gamma\in\G(|\xi|,Y)}(-1)^{t(\gamma)}=\sum_{\genfrac{}{}{0pt}{}{\xi_1+\xi_2=\xi}{|\xi_1|\cap|\xi_2|=Y}}(-1)^{\xi_1\cdot\xi_2}.
\]
\end{proposition}

Indeed, this proposition implies the equality between $Z_\lambda(G)$ and the square root $\det(I-T^\lambda(G))^{1/2}$
whose constant coefficient is equal to $+1$. It follows:
\begin{align*}
Z^I(G)&=\sum_{\xi\in Z_1(G;\Z_2)}x(|\xi|)\\
&=\sum_{\alpha\in H_1(\SI;\Z_2)}\sum_{\xi;[\xi]=\alpha}x(|\xi|) \\
&\overset{(\ref{equ:Arf})}{=}\sum_{\alpha\in H_1(\SI;\Z_2)}\Big(\frac{1}{2^g}\sum_{\lambda\in\S(\SI)}(-1)^{\Arf(\lambda)}(-1)^{q_\lambda(\alpha)}\Big)\sum_{\xi;[\xi]=\alpha}x(|\xi|)\\
&=\frac{1}{2^g}\sum_{\lambda\in \S(\SI)}(-1)^{\Arf(\lambda)}\sum_{\alpha\in H_1(\SI;\Z_2)}(-1)^{q_\lambda(\alpha)}\sum_{\xi;[\xi]=\alpha}x(|\xi|)\\
&=\frac{1}{2^g}\sum_{\lambda\in \S(\SI)}(-1)^{\Arf(\lambda)}Z_\lambda(G)\\
&=\frac{1}{2^g}\sum_{\lambda\in \S(\SI)}(-1)^{\Arf(\lambda)}\det(I-T^\lambda(G))^{1/2}.
\end{align*}

\subsection{Proof of Proposition~\ref{prop:comb}}
\label{sub:proof}

We shall now give a proof of Proposition~\ref{prop:comb}, building on and assuming results of \cite{DZMSS}.
We would like to insist on the fact that, although our proof will appear to be short, this combinatorial statement is by no mean trivial: alone, the complete proof of
a special case of Lemma~\ref{lemma:LHS} below takes 12 pages in \cite{DZMSS}, constituting the core of that paper.

We shall actually evaluate independently both sides of the equality in Proposition~\ref{prop:comb}. The right-hand side does not require too much effort.

\begin{lemma}
\label{lemma:RHS}
Fix a 1-cycle $\xi\in Z_1(G;\Z_2)$, a subgraph $Y$ of $G$ with no edge in common with $|\xi|$, and set
\[
Z(\xi,Y)=\left\{(\xi_1,\xi_2)\in Z_1(G;\Z_2)^2\,\big|\,\xi_1+\xi_2=\xi,\;|\xi_1|\cap |\xi_2|=Y\right\}.
\]
Then, the integer
\[
\sum_{(\xi_1,\xi_2)\in Z(\xi,Y)}(-1)^{\xi_1\cdot\xi_2}
\]
is equal to zero unless the following two conditions are satisfied, in which case it is equal to the cardinality of $H_1(|\xi|;\Z_2)$:
\begin{romanlist}
\item{there exists $c\in C_1(|\xi|;\Z_2)$ such that $\partial c=\partial c_Y$, where $c_Y\in C_1(G;\Z_2)$ is the 1-chain with $|c_Y|=Y$;}
\item{for any $\beta\in H_1(|\xi|;\Z_2)$ the intersection number $\beta\cdot\xi$ is equal to zero.}
\end{romanlist}
\end{lemma}
\begin{proof}
First note that any element $(\xi_1,\xi_2)$ in $Z(\xi,Y)$ provides a 1-chain $c=\xi_1+c_Y\in C_1(|\xi|;\Z_2)$ such that $\partial c=\partial c_Y$. Therefore, if there
is no such 1-chain, then the set $Z(\xi,Y)$ is empty and the integer considered equal to zero. If there is such a chain on the other hand, then the assignment
$(\xi_1,\xi_2)\mapsto \xi_1+c_Y$ defines a bijection from $Z(\xi,Y)$ onto the set of such 1-chains, with inverse $c\mapsto (c+c_Y,c+c_Y+\xi)$.
This set being an affine space over
\[
\mathrm{Ker}(\partial\colon C_1(|\xi|;\Z_2)\to C_0(|\xi|;\Z_2))=Z_1(|\xi|;\Z_2)=H_1(|\xi|;\Z_2),
\]
we get the equality
\begin{align*}
\sum_{(\xi_1,\xi_2)\in Z(\xi,Y)}(-1)^{\xi_1\cdot\xi_2}&=\sum_{\beta\in H_1(|\xi|;\Z_2)}(-1)^{(c+c_Y+\beta)\cdot(c+c_Y+\beta+\xi)}\\
&=\sum_{\beta\in H_1(|\xi|;\Z_2)}(-1)^{(c+c_Y+\beta)\cdot\xi}\\
&=\sum_{\beta\in H_1(|\xi|;\Z_2)}(-1)^{\beta\cdot\xi},
\end{align*}
as $c+c_Y$ is an element of $H_1(|\xi|;\Z_2)$. This integer vanishes unless the linear form $\beta\mapsto\beta\cdot \xi$ is identically zero, in which case it is
equal to $|H_1(|\xi|;\Z_2)|$.
\end{proof}

Proposition~\ref{prop:comb} (and therefore, Theorem~\ref{thm:KW}) now follows from one more lemma.
Once again, we shall not give here a self-contained proof of this final result, but assume statements of \cite{DZMSS}.

\begin{lemma}
\label{lemma:LHS}
Fix a 1-cycle $\xi\in Z_1(G;\Z_2)$, a subgraph $Y$ of $G$ with no edge in common with $|\xi|$, and let $\G(|\xi|,Y)$ denote the set of admissible paths in $G$
that pass once on $|\xi|$, twice on $Y$, and not on $G\setminus(|\xi|\cup Y)$. Then, the integer
\[
S(|\xi|,Y)=\sum_{\gamma\in\G(|\xi|,Y)}(-1)^{t(\gamma)}
\]
is equal to zero unless the conditions $(i)$ and $(ii)$ of Lemma~\ref{lemma:RHS} are satisfied, in which case it is equal to the cardinality of $H_1(|\xi|;\Z_2)$.
\end{lemma}
\begin{proof}
Following the terminology of~\cite[Sections 4-5]{DZMSS}, let us call an orientation of the edges of $|\xi|$ ``regular" if around each vertex $v$ of $|\xi|$,
the oriented edges belonging to $|\xi|$ that enter $v$ strictly alternate with the edges leaving it.
Observe that $|\xi|$ admits a regular orientation if and only if the condition $(ii)$ of Lemma~\ref{lemma:RHS} is satisfied.
Therefore, if condition $(ii)$ is not satisfied, then $|\xi|$ does not admit any regular orientation, and \cite[Corollary 2]{DZMSS} implies that $S(|\xi|,Y)$ is equal to zero.
On the other hand, if condition $(ii)$ is satisfied, then $|\xi|$ admits a regular orientation. Since $\Sigma$ is orientable, one easily checks that this regular orientation
can be chosen so that it extends to a ``principal orientation" on $|\xi|\cup Y$.
(This is an orientation of the edges of $|\xi|$ and of the doubled edges of $Y$ so that, around each vertex $v$ of $|\xi|\cup Y$,
the oriented edges belonging to $|\xi|\cup Y$ -- both simple and doubled -- that enter $v$ strictly alternate with the edges leaving it.)
Then, the proof of \cite[Theorem 4]{DZMSS} extends verbatim, leading to: $S(|\xi|,Y)$ vanishes unless condition $(i)$ holds, in which case $S(|\xi|,Y)=|H_1(|\xi|;\Z_2)|$.
\end{proof}


\section{The geometric proof}
\label{sec:geo}

We now give a second proof of our formula. This proof is in three steps. We first recall the definition of the dimer model, and state (an improved version of) the
Fisher correspondence \cite{Fi2} between the Ising and dimer models (Subsection~\ref{sub:Fis}). We then recall Kasteleyn's theory, as well as the geometric treatment
of the dimer model initiated by Reshetikhin and the author in \cite{C-RI} (Subsection~\ref{sub:dimers}).
Finally, we show in Subsection~\ref{sub:KWvsKF} that the $2^{2g}$ generalized Kac-Ward matrices can be naturally identified
with the $2^{2g}$ Kasteleyn matrices for the corresponding dimer model.
As a direct consequence, we obtain not only a purely geometric proof of Theorem~\ref{thm:KW}, but also the fact that both methods -- the Kac-Ward on one hand, the Fisher-Kasteleyn on the other -- should really be considered as one single method to solve the Ising model.

\subsection{Reduction of the Ising model to the dimer model}
\label{sub:Fis}

Let $\G$ be a finite graph. A {\em dimer covering\/} (or {\em perfect matching\/}) on $\G$ is a choice of edges
of $\G$, called {\em dimers\/}, such that each vertex of $\G$ is adjacent to exactly one of these edges. We shall denote by $\M(\G)$ the set of perfect matchings on $\G$.
Let us associate to each edge $e\in E(G)$ a formal variable $x_e$. The partition function for the {\em dimer model} on $\G$ is by definition
\[
Z^D(\G)=\sum_{M\in\M(\G)}x(M), \; \text{ where } \; x(M)=\prod_{e\in M}x_e. 
\]

\begin{figure}[htbp]
\labellist\small\hair 2.5pt
\pinlabel {$v$} at 190 155
\pinlabel {$G$} at 100 70
\pinlabel {$\G_G$} at 720 70
\pinlabel {$e_1$} at 270 310
\pinlabel {$e_1$} at 905 340
\pinlabel {$e_2$} at 340 215
\pinlabel {$e_2$} at 1000 210
\pinlabel {$e_3$} at 295 95
\pinlabel {$e_3$} at 925 70
\pinlabel {$e_{n-1}$} at 105 230
\pinlabel {$e_{n-1}$} at 690 235
\pinlabel {$e_n$} at 140 300
\pinlabel {$e_n$} at 740 325
\endlabellist
\centerline{\psfig{file=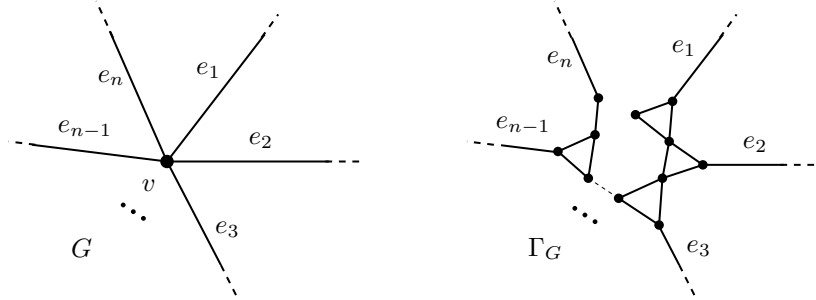,height=4cm}}
\caption{The neighborhood of a vertex $v\in V(G)$ of degree $n$, and the associated cluster in $\G_G$.}
\label{fig:blowup}
\end{figure}

As discovered by Kasteleyn \cite{Ka1} and Fisher \cite{Fi1,Fi2}, the Ising model can be interpreted as a special type of dimer model. More precisely, let $G$ be
a finite graph with variables $x_e$ associated to its edges. For each vertex $v\in V(G)$, choose a linear ordering of the adjacent edges.
Then, consider the associated graph $\G_G$ obtained from $G$ by
blowing up each vertex $v\in V(G)$ into a {\em cluster}, as illustrated in Figure~\ref{fig:blowup}.

In this way, each vertex $v\in V(G)$ of degree $n$ gives rise to $2n$ vertices in $\G_G$. Furthermore, there is a natural inclusion of $E(G)$ into $E(\G_G)$.
To each edge of $\G_G$ coming from an edge $e$ of $G$, assign the same variable $x_e$. To each other (new) edge of $\G_G$, assign the weight $1$.
Obviously, the graph $\G_G$ depends on the choice of the linear orderings of the edges around each vertex. However, for any such choice, we have the following result.

\begin{proposition}
\label{prop:Fis}
The restriction map
\[
\varphi\colon C_1(\G_G;\Z_2)\to C_1(G;\Z_2),\; \sum_{e\in E(\G_G)}\xi_e\,e\mapsto \sum_{e\in E(G)}\xi_e\,e
\]
induces a bijection $\M(\G_G)\to Z_1(G;\Z_2)$ so that $x(|\varphi(M)|)=x(M)$ for any perfect matching $M\in\M(\G_G)$. In particular, $Z^I(G)=Z^D(\G_G)$.
\end{proposition}
\begin{proof}
First note that for any $M\in\M(\G_G)$, $\varphi(M)$ is a 1-cycle modulo 2. Indeed, let $v\in V(G)$ be a fixed vertex and let $n$ denote its degree.
In and around the cluster coming from $v$, $M$ matches exactly $2n+\delta_M(v)$ vertices, where $\delta_M(v)$ denotes the degree of $v$ in the subgraph $|\varphi(M)|\subset G$.
As $M$ is a matching, it matches an even number of vertices, so $\delta_M(v)$ is even and $\varphi(M)$ is a 1-cycle modulo 2.
\parpic[r]{$\begin{array}{c}
\includegraphics[height=2.5cm]{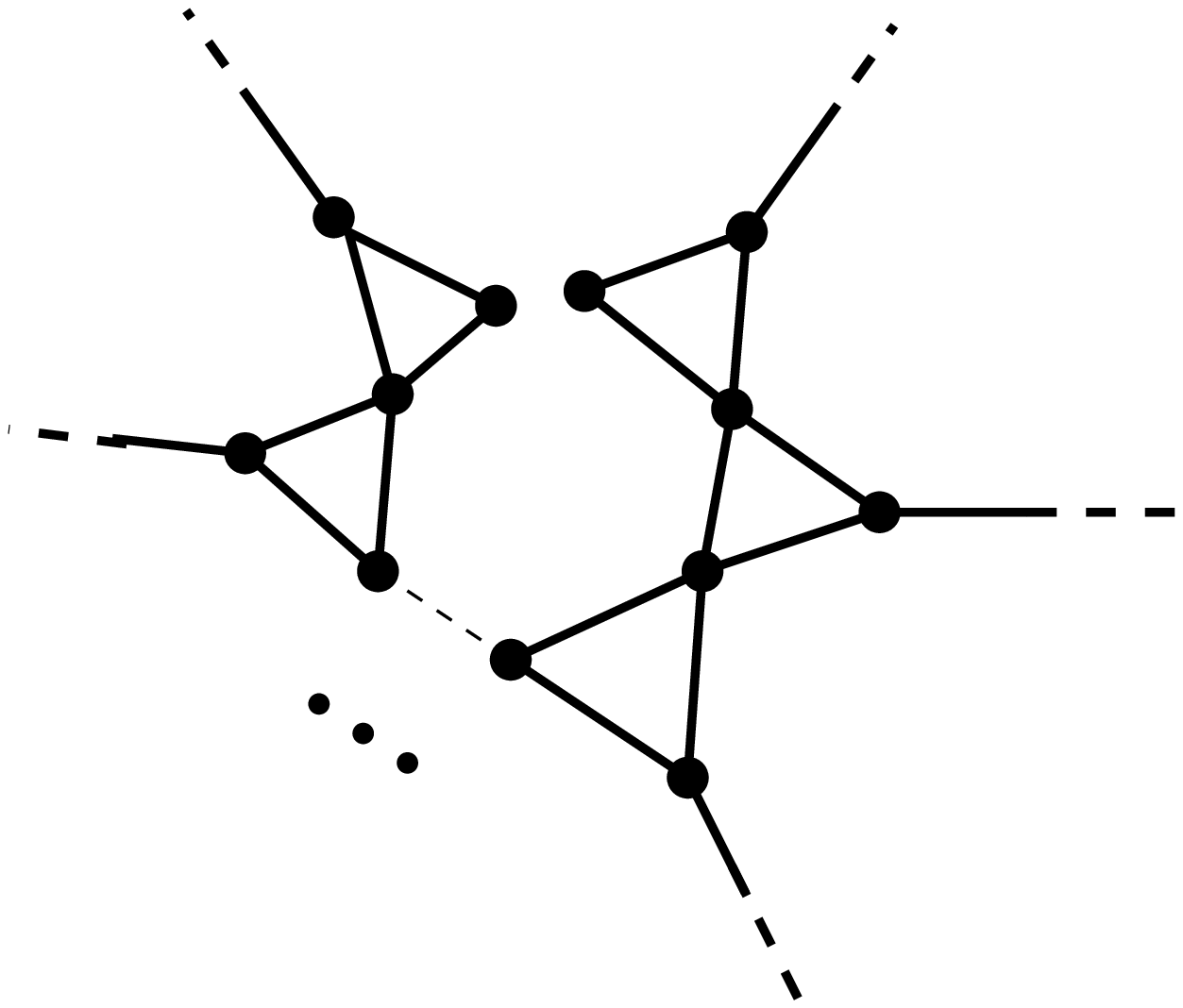}
\end{array}$}
Then, one easily checks by induction on the degree of $v$ that any $\xi\in C_1(G;\Z_2)$ with $\partial\xi$ even at $v$ (resp. odd at $v$) uniquely extends to a perfect
matching on the cluster corresponding to $v$ (resp. on the graph illustrated opposite). In particular, any $\xi\in Z_1(G;\Z_2)$ extends uniquely to a perfect
matching $M_\xi\in\M(\G_G)$. By construction, $\varphi(M_\xi)=\xi$ for any $\xi\in Z_1(G;\Z_2)$, so $\varphi$ induces a bijection as claimed.
The equality $x(|\varphi(M)|)=x(M)$ follows from the definitions, and the proposition is proved.
\end{proof}

\begin{figure}[htbp]
\labellist\small\hair 2.5pt
\pinlabel {$M_0$} at 40 55
\endlabellist
\centerline{\psfig{file=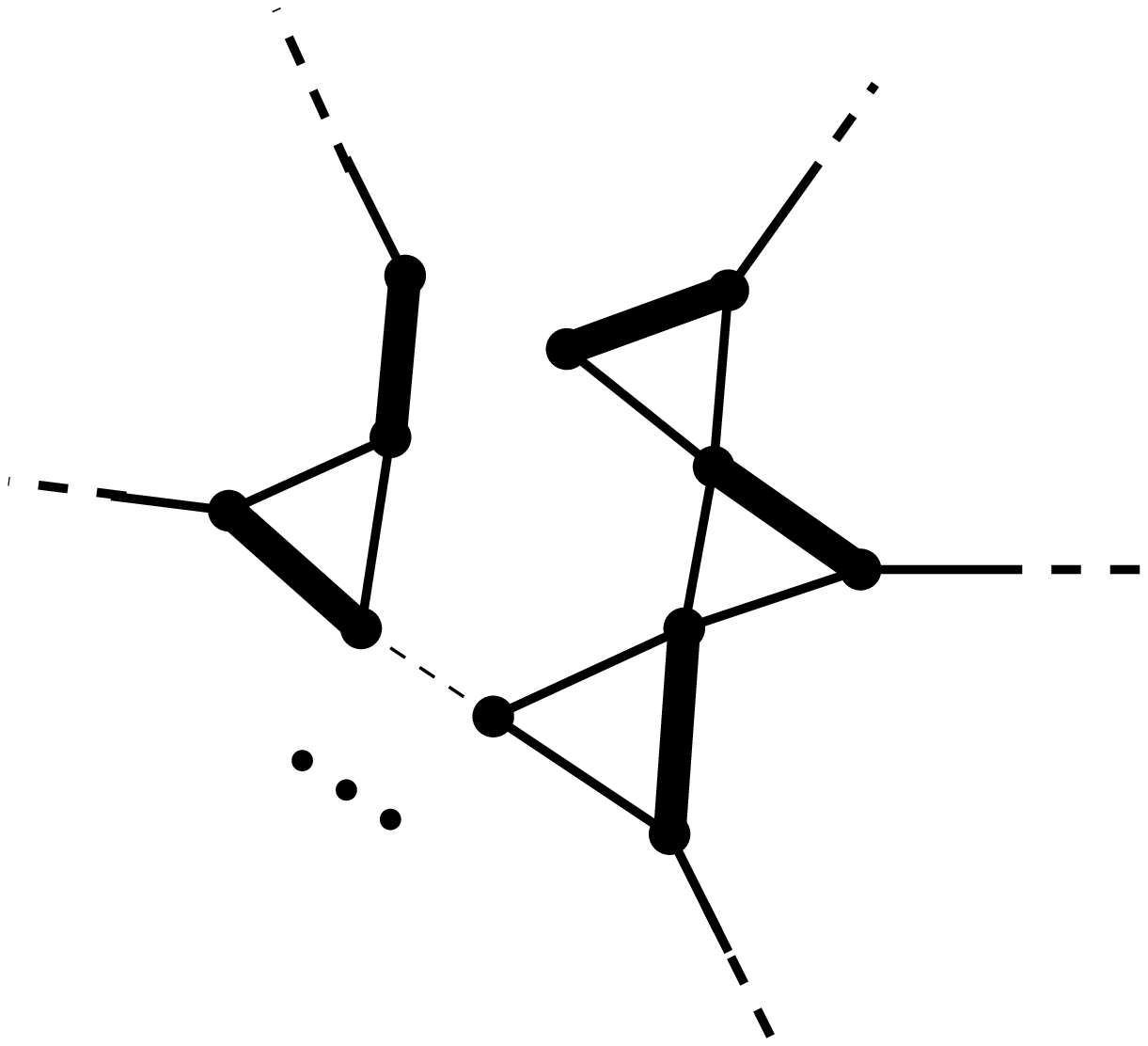,height=3cm}}
\caption{The preferred perfect matching $M_0$ on $\G_G$.}
\label{fig:M0}
\end{figure}
In particular, there is a preferred perfect matching $M_0$ on $\G_G$, namely the unique one such that $\varphi(M_0)=0$. It is illustrated in Figure~\ref{fig:M0}.

Note also that if the first graph $G$ is embedded in a surface $\SI$, and if the chosen linear orderings around the vertices are compatible with the cyclic orderings induced by the embedding $G\subset\SI$, then the graph $\G_G$ also embeds in the same surface $\SI$. Furthermore, the bijection $\varphi$ can easily be seen to satisfy the following additional property: for any $M\in\M(\G_G)$, the 1-cycles $M+M_0\in Z_1(\G_G;\Z_2)$ and
$\varphi(M)\in Z_1(G;\Z_2)$ define the same homology class in $H_1(\SI;\Z_2)$. Therefore, for any $\alpha\in H_1(\SI;\Z_2)$, we have equality between the corresponding
partial partition functions for the Ising and dimer models:
\begin{equation}
\label{equ:ref}
Z_\alpha^I(G):=\sum_{\genfrac{}{}{0pt}{}{\xi\in Z_1(G;\Z_2)}{[\xi]=\alpha}}x(|\xi|)=\sum_{\genfrac{}{}{0pt}{}{M\in\M(\G_G)}{[M+M_0]=\alpha}}x(M)=:Z_\alpha^D(\G_G).
\end{equation}

\begin{remark}
The construction described here is not the one given by Fisher in~\cite{Fi2}. In his original construction, each vertex of degree $n$ gives rise to $3n-6$ vertices.
It is therefore simpler for vertices of degree $n\le 5$, in particular for square lattices. However, the construction presented here is simpler
for vertices of degree $n\ge 7$, and overall more suited for the purpose of this paper.
\end{remark}

\subsection{The geometric treatment of the dimer model}
\label{sub:dimers}

In this subsection, we recall the main features of the geometric method initiated in~\cite{C-RI}, following the simplified version given in \cite{Cim}.
We refer to these articles for further details, and to~\cite{D96,Tes,G-L} for earlier treatments of the same problem.

\subsubsection{Dimers and Pfaffians: Kasteleyn theory}
\label{subsub:Kast}

Let $\G$ be a finite graph. The aim is to compute the dimer partition function
\[
Z^D(\G)=\sum_{M\in\M(\G)}x(M), \; \text{ where } \; x(M)=\prod_{e\in M}x_e.
\]
Kasteleyn's method is based on the following beautifully simple computation. If there exists a perfect matching,
then the number of vertices of $\G$ is even. Enumerate them by $1,2,\dots,2n$, and fix an arbitrary orientation $K$ of the edges
of $\G$.

\begin{definition}
\label{def:Kast}
The associated {\em Kasteleyn matrix\/} $A^K(\G)=(A_{ij}^K)$ is the $2n\times 2n$ skew-symmetric matrix whose coefficients are given by
\[
A^K_{ij}=\sum_{e}\e_{ij}^K(e)x_e,
\]
where the sum is over all edges $e$ in $\Gamma$ between the vertices $i$ and $j$, and
\[
\e^K_{ij}(e)=
\begin{cases}
\phantom{-}1 & \text{if $e$ is oriented by $K$ from $i$ to $j$;} \\
-1 & \text{otherwise.}
\end{cases}
\]
\end{definition}

Recall that the Pfaffian of a skew-symmetric matrix $A=(A_{ij})$ of size $2n$ is given by
\[
\Pf(A)=\sum_{[\sigma]\in\Pi}(-1)^{\text{sgn}(\sigma)} A_{\sigma(1)\sigma(2)}\cdots A_{\sigma(2n-1)\sigma(2n)},
\]
where the sum is over the set $\Pi$ of perfect matchings of $\{1,\dots,2n\}$ and $\sigma$ is a permutation of $\{1,\dots,2n\}$
representing the matching $[\sigma]$. In the case of $A^K(\G)$,
a matching of $\{1,\dots,2n\}$ contributes to the Pfaffian if and only if it is realized by a perfect matching
on $\G$, and this contribution is $\pm x(M)$. More precisely,
\begin{equation}
\label{equ:Pf}
\Pf(A^K(\G))=\sum_{M\in\M(\G)}\e^K(M)x(M),
\end{equation}
where the sign $\e^K(M)$ can be computed as follows: if the perfect matching $M$ is given by edges $e_1,\dots,e_n$ matching vertices $i_\ell$ and $j_\ell$ for $\ell=1,\dots,n$, let $\sigma$ denote the permutation sending
$(1,\dots, 2n)$ to $(i_1,j_1,\dots,i_n,j_n)$, and set
\begin{equation}
\label{equ:eps}
\e^K(M)=(-1)^{\text{sgn}(\sigma)}\prod_{\ell=1}^n\e^K_{i_\ell j_\ell}(e_\ell).
\end{equation}
The problem of expressing $Z^D(G)$ as a Pfaffian now boils down to finding an orientation $K$ of the edges of $\G$
such that $\e^K(M)$ does not depend on $M$.

Obviously, any perfect matching $M$ can be considered as a cellular 1-chain $M\in C_1(\G;\Z_2)$ such that
$\partial M=\sum_{v}v$, the sum being on all vertices of $\G$.
Hence, given any two perfect matchings $M,M'$, their sum $M+M'$ is a 1-cycle.
The connected components of this 1-cycle are disjoint simple loops of even length; let us denote them by
$\{C_i\}_i$. An easy computation shows that
\begin{equation}
\label{equ:cc}
\e^K(M)\e^K(M')=\prod_i(-1)^{n^K(C_i)+1},
\end{equation}
where $n^K(C_i)$ denotes the number of edges of $C_i$ where a fixed orientation of $C_i$ differs from $K$.
(Since $C_i$ has even length, the parity of this number is independent of the orientation of $C_i$.)
Therefore, we are now left with the problem of finding an orientation $K$ of $\G$ such that, for any cycle $C$ of even
length such that $\G\setminus C$ admits a perfect matching, $n^K(C)$ is odd. Such an orientation is called a {\em Pfaffian orientation\/}.
By the discussion above, if $K$ is a Pfaffian orientation, then $Z^D(\G)=|\Pf(A^K(\G))|$.

Kasteleyn's early triumph was to prove that every planar graph admits a Pfaffian orientation. More precisely, let
$\G$ be a graph embedded in the plane. Each face $f$ of $\G\subset\R^2$ inherits the (say, counterclockwise)
orientation of $\R^2$, so $\partial f$ can be oriented as the boundary of the oriented face $f$.

\begin{Kast}[\cite{Ka2,Ka3}]
Given $\G\subset\R^2$, there exists an orientation $K$ of $\G$ such that, for each face $f$
of $\G\subset\R^2$, $n^K(\partial f)$ is odd. Furthermore, such an orientation is Pfaffian.
\end{Kast}

An amazing consequence of this result is that it enables to compute the partition function of the dimer model
on a planar graph in polynomial time. There is no hope to extend this result to the general case. Indeed, some graphs (such as the complete bipartite graph
$K_{3,3}$) do not admit a Pfaffian orientation. More generally, enumerating the perfect matchings on a (bipartite) graph
is a $\#P$-complete problem \cite{Val}. It turns out that Kasteleyn's method does extend to surfaces, but one needs to compute
many Pfaffians. This is the aim of the following two paragraphs.

\subsubsection{Kasteleyn orientations as discrete spin structures}
\label{subsub:spin}

First note that any finite connected graph $\G$ can be embedded in a closed oriented connected surface $\SI$ as the 1-skeleton of a cellular decomposition $X$
of $\SI$. (This simply means that the complement of $\G$ in $\SI$ consists of open 2-discs.)
In this paragraph, we shall use the same notation $X$ for the graph embedded in the surface and for the induced cell complex.

An orientation $K$ of the 1-cells of such an embedded graph is called a {\em Kasteleyn orientation on $X$\/} if,
for each 2-cell $f$ of $X$, the following condition holds: the number $n^K(\partial f)$ of edges in $\partial f$
where $K$ disagrees with the orientation on $\partial f$ induced by the counterclockwise orientation on $f$, is odd.
Given a Kasteleyn orientation on $X$, there is an obvious way to obtain another one: pick a vertex of $X$ and flip
the orientation of all the edges adjacent to it. Two Kasteleyn orientations are said to be {\em equivalent\/} if
they can be related by such moves. Let us denote by $\K(X)$ the set of equivalence classes of Kasteleyn
orientations on $X$. 

\begin{proposition}
\label{prop:Kast}
A embedded graph $X$ admits a Kasteleyn orientation if and only if $X$ has an even number of vertices.
In this case, the set $\K(X)$ is an affine $H^1(\SI;\Z_2)$-space.
\end{proposition}

The easy proof can be found in \cite[Section 4]{C-RI}.

Now, the game consists in trying to encode combinatorially a spin structure on a surface $\SI$, or
equivalently, a vector field on $\SI$ with isolated zeroes of even index. Let us begin by fixing a cellular decomposition
$X$ of $\SI$.

$\bullet$
To construct a (unit length) vector field along the 0-skeleton $X^0$, we just need to specify one tangent
direction at each vertex of $X$. Such an information is given by a perfect matching $M$ on $X^1$: at each vertex,
point in the direction of the adjacent dimer.

\begin{figure}[htbp]
\labellist\small\hair 2.5pt
\pinlabel {$K$} at 220 165
\pinlabel {$M$} at 75 60
\pinlabel {$M$} at 395 190
\endlabellist
\centerline{\psfig{file=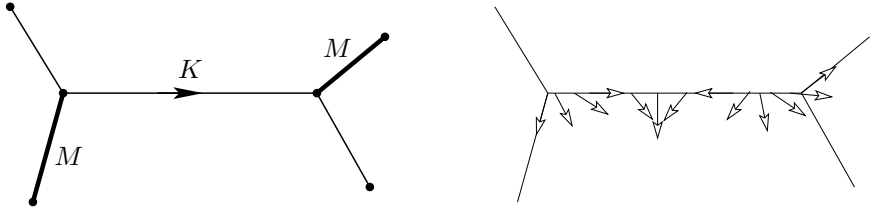,height=2.7cm}}
\caption{Construction of the vector field along the 1-skeleton of $X$.}
\label{fig:vector}
\end{figure}

$\bullet$
This vector field along $X^0$ extends to a unit vector field on $X^1$, but not uniquely. Roughly speaking,
it extends in two different natural ways along each edge of $X^1$, depending on the sense of rotation of the resulting
vector field. We shall
encode this choice by an orientation $K$ of the edges of $X^1$, together with the following convention: moving along
an oriented edge, the tangent vector first rotates counterclockwise until it points in the direction of the edge,
then rotates clockwise until it points backwards, and finally rotates counterclockwise until it coincides with
the tangent vector at the end vertex. This is illustrated in Figure~\ref{fig:vector}.

$\bullet$
Each face of $X$ being homeomorphic to a 2-disc, the unit vector field defined along $X^1$ naturally extends to
a vector field $\lambda_M^K$ on $X$, with one isolated zero in the interior of each face. One easily checks that
for each face $f$ of $X$, the index of the zero of $\lambda_M^K$ in $f$ has the parity of $n^K(\partial f)+1$ (see \cite[Lemma 4.1]{Cim} for the proof).
Therefore, this vector field defines a spin structure if and only if $K$ is a Kasteleyn orientation.

To summarize, a perfect matching $M$ on $X^1$ and a Kasteleyn orientation $K$ on $X$ determine a spin
structure on $\SI$. By Johnson's theorem (recall Subsection~\ref{sub:Joh}), it automatically defines a quadratic form $q^K_M\colon H_1(\SI;\Z_2)\to\Z_2$.
This form is characterized by the following property: if $C$ is an oriented simple closed curve on $X^1$, then
\begin{equation}
\label{equ:quadratic}
q^K_M([C])=n^K(C)+\ell_M(C)+1\pmod{2},
\end{equation}
where $\ell_M(C)$ denotes the number of vertices in $C$ such that the adjacent dimer of $M$ points out to the left of $C$. (See~\cite[Lemma 3.3]{Cim}.)
This fact easily implies the following correspondence theorem.

\begin{theorem}
\label{thm:corr}
Let $X$ be a cellular decomposition of an oriented closed surface $\SI$. Then, any perfect matching $M\in\M(X^1)$
induces an $H^1(\SI;\Z_2)$-equivariant bijection
\[
\psi_M\colon\K(X)\to\Q(\SI)=\S(\SI),\quad [K]\mapsto q_M^K
\]
from the set of equivalence classes of Kasteleyn orientations on $X$ onto the set of spin structures on $\SI$. \qed
\end{theorem}

\subsubsection{The Pfaffian formula}
\label{subsub:Pfaff}

As a direct consequence of Equation~(\ref{equ:quadratic}), we obtain ``for free" the following non-trivial combinatorial result.

\begin{proposition}
\label{prop:quadratic}
Let $K$ be a Kasteleyn orientation on $X$, and $M$ be a perfect matching on $X^1$. Given a homology class
$\alpha\in H_1(\SI;\Z_2)$, represent it by oriented simple closed curves $C_1,\dots,C_m$ in $X^1$. Then, the equality
\[
q_M^K(\alpha)=\sum_{i=1}^m(n^K(C_i)+\ell_M(C_i)+1)+\sum_{1\le i<j\le m}C_i\cdot C_j\pmod{2}
\]
determines a well-defined quadratic form $q_M^K\colon H_1(\SI;\Z_2)\to\Z_2$.\qed
\end{proposition}

We shall now use this combinatorial information, together with the results and notation of Paragraph~\ref{subsub:Kast},
to derive our Pfaffian formula.

Let $\G$ be a finite connected graph. If $\G$ does not admit any perfect matching, then the partition function $Z^D(\G)$ is obviously zero. So,
let us assume that $\G$ admits a perfect matching $M_0$. Enumerate the vertices of $\G$ by $1,2,\dots,2n$ and embed $\G$ in a closed orientable surface
$\SI$ of genus $g$ as the 1-skeleton of a cellular decomposition $X$ of $\SI$.

Since $\G$ has an even number of vertices, the set $\K(X)$ is an affine $H^1(\SI;\Z_2)$-space (Proposition~\ref{prop:Kast}). For any Kasteleyn orientation
$K$, the Pfaffian of the associated weighted skew-adjacency matrix satisfies
\begin{align*}
\e^K(M_0)\Pf(A^K(\G))&\overset{(\ref{equ:Pf})}{=}\sum_{M\in\M(\G)}\e^K(M_0)\e^K(M)\,x(M)\\
	&\overset{(\ref{equ:cc})}{=}\sum_{M\in\M(\G)}(-1)^{\sum_i (n^K(C_i)+1)}x(M),
\end{align*}
where the $C_i$'s are the connected components of the cycle $M+M_0\in C_1(X;\Z_2)$.
Note that given any vertex of $C_i$, the adjacent dimer of $M_0$ lies on $C_i$, so that $\ell_{M_0}(C_i)=0$.
Since the cycles $C_i$ are disjoint, Proposition~\ref{prop:quadratic} gives
\[
\sum_i(n^K(C_i)+1)=\sum_i(n^K(C_i)+\ell_{M_0}(C_i)+1)=q^K_{M_0}([M+M_0]).
\]
Therefore, every element $[K]$ of $\K(X)$ induces a linear equation
\begin{equation}
\label{equ:Pf'}
\e^K(M_0)\Pf(A^K(\G))=\sum_{\alpha\in H_1(\SI;\Z_2)}(-1)^{q^K_{M_0}(\alpha)}Z^D_{\alpha}(\G),
\end{equation}
where $Z^D_{\alpha}(\G)=\sum_{[M+M_0]=\alpha}x(M)$, the sum being over all $M\in\M(\G)$ such that $[M+M_0]=\alpha$.
Equation~(\ref{equ:Arf}) now immediately yields the following Pfaffian formula.

\begin{theorem}
\label{thm:Pf}
Let $\G$ be a graph embedded in a closed oriented surface $\SI$ of genus $g$ such that $\SI\setminus\G$ consists of
open 2-discs. Then, the partition function of the dimer model on $\Gamma$ is given by the formula
\[
Z^D(\G)=\frac{1}{2^{g}}\sum_{[K]\in\K(X)}(-1)^{\Arf(q^{K}_{M_0})}\e^K(M_0)\Pf(A^{K}(\G)),
\]
where the sum is taken over all equivalence classes of Kasteleyn
orientations, and $\Arf(q)\in\Z_2$ denotes the Arf invariant of the quadratic form $q$.\qed
\end{theorem}

\subsection{Identification of the Kac-Ward and Fisher-Kasteleyn matrices}
\label{sub:KWvsKF}

We are now ready to put together the results of the two previous subsections to obtain the geometric proof of Theorem~\ref{thm:KW}.

\begin{figure}[htbp]
\labellist\small\hair 2.5pt
\pinlabel {$a_1$} at 250 220
\pinlabel {$a_2$} at 282 157
\pinlabel {$a_3$} at 237 53
\pinlabel {$b_1$} at 170 232
\pinlabel {$b_2$} at 193 167
\pinlabel {$b_3$} at 187 135
\pinlabel {$a_{n-1}$} at 46 133
\pinlabel {$a_n$} at 95 227
\pinlabel {$b_n$} at 145 175
\endlabellist
\centerline{\psfig{file=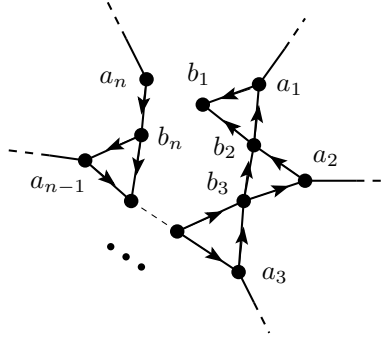,height=4.5cm}}
\caption{The numbering of the vertices, and a Kasteleyn orientation at a cluster.}
\label{fig:K}
\end{figure}

Let $G$ be a finite graph. First, number the vertices of $G$. Then, embed $G$ in a closed oriented surface $\SI$, and blow up each vertex $v\in V(G)$ into a cluster
as described in Figure~\ref{fig:blowup}. This gives another graph $\G_G$, also embedded in $\SI$. Enumerate the vertices of $\G_G$ in each cluster by
$\{a_1,b_1,\dots,a_n,b_n\}$, as illustrated in Figure~\ref{fig:K}. This, together with the numbering of the vertices of $G$, defines an ordering of the vertices of $\G_G$.

Let $K$ be any Kasteleyn orientation on $\G_G\subset\SI$. Replacing $K$ by an equivalent Kasteleyn orientation, it may be assumed that it orients the edges in each
cluster as described in Figure~\ref{fig:K}. Let $A^K(\G_G)$ denote the associated Kasteleyn matrix (recall Definition~\ref{def:Kast}).

Also, let $\lambda_{M_0}^K$ denote the vector field along $\G_G$ induced by $K$ and by the preferred perfect matching $M_0$ (Figure~\ref{fig:M0}), as
described in Paragraph~\ref{subsub:spin}. This vector field extends to a nowhere vanishing vector field in the neighborhood of each cluster, thus defining a vector field
$\lambda$ along $G$. Let $I-T^\lambda(G)$ be the associated generalized Kac-Ward matrix (recall Definition~\ref{def:KW}).

\begin{proposition}
\label{prop:KWKF}
We have the equality $\Pf(A^K(\G_G))=\det(I-T^\lambda(G))^{1/2}$.
\end{proposition}


Note that with our enumeration convention, the sign $\e^K(M_0)$ is equal to $+1$ (recall Equation~(\ref{equ:eps})). Therefore, we immediately obtain Theorem~\ref{thm:KW}:
\begin{align*}
Z^I(G)&\overset{\ref{prop:Fis}}{=}Z^D(\G_G)\\
&\overset{\ref{thm:Pf}}{=}\frac{1}{2^{g}}\sum_{[K]\in\K(X)}(-1)^{\Arf(q^{K}_{M_0})}\Pf(A^{K}(\G_G))\\
&\overset{\ref{thm:corr}}{=}\frac{1}{2^{g}}\sum_{\lambda\in \S(\SI)}(-1)^{\Arf(\lambda)}\Pf(A^{K}(\G_G))\\
&\overset{\ref{prop:KWKF}}{=}\frac{1}{2^g}\sum_{\lambda\in \S(\SI)}(-1)^{\Arf(\lambda)}\det(I-T^\lambda(G))^{1/2}.
\end{align*}

\begin{proof}[Proof of Proposition~\ref{prop:KWKF}]
First note that the constant coefficient of the polynomial $\det(I-T^\lambda(G))^{1/2}$ is (by choice of the square root) equal to $+1$.
On the other hand, the constant coefficient
of $\Pf(A^K(\G_G))$ is equal to $\prod_{v\in V(G)}\Pf(A_{\text{deg}(v)})$, where $A_n$ is the $2n\times 2n$ Kasteleyn matrix associated to a cluster coming from a vertex
$v\in V(G)$ of degree $n$. One checks by direct computation that $\Pf(A_n)$ is equal to $+1$ for any $n$. Hence, both polynomials in the statement of
Proposition~\ref{prop:KWKF} have the same constant coefficient. It is therefore enough to show that their squares coincide, that is, to check the equality
\[
\det(A^K(\G_G))=\det(I-T^\lambda(G)).\tag{$\star$}
\]

Then, observe that both sides of this equation are left unchanged when adding an edge $e$ to $G\subset\SI$ with associated weight $x_e=0$. Indeed, if $G'\subset\SI$
denotes this new graph, then $I-T^\lambda(G')=(I-T^\lambda(G))\oplus I_2$, where $I_2$ denotes the size 2 identity matrix. As the left-hand side of $(\star)$ is equal to
\[
\det(A^K(\G_G))\overset{(\ref{equ:Pf'})}{=}\Big(\sum_\alpha (-1)^{q^K_{M_0}(\alpha)} Z_\alpha^D(\G_G)\Big)^2\overset{(\ref{equ:ref})}{=}
\Big(\sum_\alpha (-1)^{q^K_{M_0}(\alpha)} Z_\alpha^I(G)\Big)^2,
\]
it is obviously also left unchanged. Therefore, it may be assumed without loss of generality that all vertices $v\in V(G)$ have even degree. (This assumption is by no mean crucial, but it will simplify the computations to come.) Indeed, given two vertices $v,v'\in V(G)$ of odd degree, double each edge in a path from $v$ to $v'$. The new graph
will now have even degree at both $v$ and $v'$, while all the other degree parities are unchanged. As the sum $\sum_{v\in V(G)}{\text{deg}}(v)=2|E(G)|$ is even, the number
of vertices in $G$ with odd degree is even, and the procedure above allows to get rid of all of them.

Next, it may be assumed that $G$ has no loop. (Again, this is only for notational simplicity in the computations to come.) Indeed, given a loop $e$ at a vertex $v\in V(G)$,
add a vertex $v'$ in it: this divides $e$ into two edges to which we assign the weights $x_e$ and $1$. Let $G''\subset\SI$ denote this new graph.
It is an amusing exercise to check that the right-hand side of $(\star)$ is unchanged when replacing $G$ by $G''$: details are left to the reader. As $Z^I_\alpha(G)$ and
$Z^I_\alpha(G'')$ obviously coincide, the invariance of the left-hand side of $(\star)$ follows directly from the equation displayed above.

So, let $G$ be a finite graph with no loops, and with each vertex of even degree. The strategy to prove $(\star)$ will be to transform $A^K(\G_G)$ into
the matrix $I-T^\lambda(G)$ using moves that do not change the determinant.
Let us focus on an arbitrary vertex $v\in V(G)$ of degree $n$. By definition of $K$ and numbering convention of the vertices in the corresponding
cluster (recall Figure~\ref{fig:K}), and since there is no loop at $v$, the corresponding part of the matrix $A^K(\G_G)$ has the following form
\[
A^K(\G_G)=\left(
\begin{array}{ccccccccccccc}
\ast & \phantom{-}c_1 & 0 & \phantom{-}c_2 & \phantom{-}0 & \phantom{-}c_3 & \phantom{-}0 & \dots & 0& \phantom{-}c_n & \phantom{-}0 & \ast \cr
r_1  & \phantom{-}0   & 1 & \phantom{-}0   & -1&     &   &       &  &     &   & r'_1 \cr
  0  & -1  & 0 & \phantom{-}0   & -1&     &   &       &  &     &   & 0 \cr
r_2  & \phantom{-}0   & 0 & \phantom{-}0   & \phantom{-}1 &  \phantom{-}0  & -1&       &  &     &   & r'_2 \cr
  0  & \phantom{-}1   & 1 & -1  & \phantom{-}0 &  \phantom{-}0  & -1&       &  &     &   & 0 \cr
r_3  &     &   & \phantom{-}0   & \phantom{-}0 &  \phantom{-}0  &  \phantom{-}1&       &  &     &   & r'_3 \cr
  0  &     &   & \phantom{-}1  & \phantom{-}1 & -1  &  \phantom{-}0&       &  &     &   & 0 \cr
\vdots&     &   &     &   &     &   & \ddots   & &   \phantom{-}0 & -1& \vdots \cr
 0   &     &   &     &   &     &   &       &  &   \phantom{-}0 & -1& 0 \cr
r_n  &     &   &     &   &     &   &     0 & 0&   \phantom{-}0 & \phantom{-}1 & r'_n \cr
  0  &     &   &     &   &     &   &     1 & 1&  -1 & \phantom{-}0 & 0 \cr
\ast & \phantom{-}c'_1 & 0 & \phantom{-}c'_2 & 0 & c'_3 & \phantom{-}0 & \dots & 0& c'_n & \phantom{-}0 & \ast 
\end{array}\right),
\]
where $r_i,r_i'$ (resp. $c_j,c_j'$) denote rows (resp. columns) of the matrix. Using elementary row and column operations and the fact that $n$ is even,
one can easily modify $A^K(\G_G)$ to obtain
\[
A^K(\G'_G)=
\begin{pmatrix}
\ast&\phantom{-}c_1&\phantom{-}c_2&\phantom{-}c_3&\phantom{-}c_4&\dots&\phantom{-}c_n&\ast\\
r_1&\phantom{-}0&\phantom{-}1&-1&\phantom{-}1&\dots&\phantom{-}1&r_1'\\
r_2&-1&\phantom{-}0&\phantom{-}1&-1&\dots&-1&r_2'\\
r_3&\phantom{-}1&-1&\phantom{-}0&\phantom{-}1&\dots&\phantom{-}1&r_3'\\
r_4&-1&\phantom{-}1&-1&\phantom{-}0&\dots&-1&r_4'\\
\vdots&\phantom{-}\vdots & & &\phantom{-}\vdots&\ddots&\phantom{-}\vdots&\vdots\\
r_n&-1&\phantom{-}1&-1&\phantom{-}1&\dots&\phantom{-}0&r'_n\\
\ast&\phantom{-}c'_1&\phantom{-}c'_2&\phantom{-}c'_3&\phantom{-}c'_4&\dots&\phantom{-}c'_n&\ast
\end{pmatrix}.
\]
It is of historical interest to note that this is precisely the type of matrices first considered by Kasteleyn~\cite{Ka1} and Fisher~\cite{Fi1} to solve the Ising model
on square lattices (case $n=4$). It is not strictly speaking a Kasteleyn matrix, as the corresponding graph $\G'_G$ does not embed in the same surface as $G$ anymore:
every vertex $v\in V(G)$ of degree $n$ gives rise to a complete graph on $n$ vertices. However, its Pfaffian still gives the right answer for the Ising model on $G$.
It is only later that Fisher~\cite{Fi2} found his correspondence $G\mapsto\G_G$ where $\G_G$ remains embedded in the same surface as $G$.

Let $L_n$ denote the $n\times n$ skew-symmetric matrix appearing above. Its determinant is equal to $1$, and its inverse is given by
\[
L_n^{-1}=
\begin{pmatrix}
0&-1&-1&\dots&-1\\
1&\phantom{-}0&-1&\dots&-1\\
1&\phantom{-}1&\phantom{-}0&\dots&-1\\
\vdots&\phantom{-}\vdots&\phantom{-}\vdots&\ddots&\phantom{-}\vdots\\
1&\phantom{-}1&\phantom{-}1&\dots&\phantom{-}0
\end{pmatrix}.
\]
Therefore,
\[
\det(A^K(\G_G))=\det(A^K(\G'_G))=\det\Big(A^K(\G'_G)\cdot\bigoplus_{v\in V(G)}L^{-1}_{\text{deg}(v)}\Big).
\]
Let $\widetilde{B}^K$ denote this latter matrix. Its rows and columns are indexed by $V(\G'_G)=\bigcup_{v\in V(G)} C(v)\subset V(\G_G)$, where $C(v)$ denotes the set of
vertices $\{a_1,a_2\dots,a_{\text{deg}(v)}\}$ in the cluster associated to $v\in V(G)$ (recall Figure~\ref{fig:K}). The coefficient of $\widetilde{B}^K$ corresponding to
$a\in C(v)$ and $a'\in C(v')$ is equal to
\[
\widetilde{B}^K_{a,a'}=\begin{cases}
1& \text{if $a=a'$,} \\ 
\e_{aa''}^K(e)\ell_{a''a'}\,x_e& \text{if there is an $a''\in C(v')$ joined to $a$ in $\G_G$,} \\
0 & \text{otherwise,}
\end{cases}
\]
where $e\in E(\G_G)$ denotes the edge joining $a$ and $a''$, $\e_{aa''}^K(e)=1$ if $e$ is oriented by $K$ from $a$ to $a''$, $\e_{aa'}^K(e)=-1$ otherwise, and
\[
\ell_{a''a'}=\begin{cases}
+1& \text{if $a'<a''$ in the linear ordering of the vertices of $C(v')$,} \\ 
-1& \text{if $a'>a''$,} \\
\phantom{-}0 & \text{if $a'=a''$.}
\end{cases}
\]
Finally, let $B^K$ denote the matrix obtained from $\widetilde{B}^K$ by multiplying by $i=\sqrt{-1}$ (resp. by $-i$) each row (resp. each column) corresponding to a vertex
$a\in V(\G_G)$ whose adjacent edge in $E(G)\subset E(\G_G)$ is oriented by $K$ towards $a$. We shall write $K(a)=1$ if this condition holds, and $K(a)=0$ otherwise, so that
$\e_{aa''}^K(e)=(-1)^{K(a)}$. Obviously,
\[
\det(B^K)=\det(\widetilde{B}^K)=\det(A^K(\G_G)),
\]
and the coefficient of $B^K$ corresponding to $a\in C(v)$ and $a'\in C(v')$ is
\[
B^K_{a,a'}=\begin{cases}
1& \text{if $a=a'$,} \\ 
(-i)^{K(a)+K(a')}\ell_{a''a'}\,x_e& \text{if there is an $a''\in C(v')$ joined to $a$ in $\G_G$,} \\
0 & \text{otherwise.}
\end{cases}
\]
Now, recall that the matrix $I-T^\lambda(G)$ is indexed by the set $E$ of oriented edges of $G$. Let $\varphi\colon E\to V(\G_G')$ denote the bijection mapping each oriented
edge $e$ to its starting vertex in $\G_G$. By definition of the matrix $T^\lambda(G)$, and since $G$ has no loops,
\[
(I-T^\lambda(G))_{e,e'}=\begin{cases}
1& \text{if $e=e'$,} \\ 
-\exp\left(\frac{i}{2}\alpha_\lambda(e,e')\right)\,x_e& \text{if $f(e)=s(e')$ but $e'\neq -e$,} \\
0 & \text{otherwise.}
\end{cases}
\]
For the two oriented edges $e,e'$ of $G$, let $s(e)=v$, $s(e')=v'$ denote their starting vertices in $G$, and $\varphi(e)=a$, $\varphi(e')=a'$ their starting vertices in $\G_G$.
Then, the condition $f(e)=s(e')$ and $e'\neq -e$ is equivalent to the condition: the oriented edge $e$ of $\G_G$ joins $a$ to some $a''\in C(v')$ with $a''\neq a'$.
Therefore, the coefficient $(I-T^\lambda(G))_{e,e'}$ is non-zero if and only if $B^K_{\varphi(e),\varphi(e')}$ is non-zero.
Finally, using the construction of the vector field $\lambda=\lambda_{M_0}^K$ induced by $K$ and $M_0$, one easily checks that for any $e,e'\in E$ satisfying this condition,
\[
\textstyle-\exp\left(\frac{i}{2}\alpha_\lambda(e,e')\right)=(-i)^{K(a)+K(a')}\ell_{a''a'}.
\]
Therefore, the matrices $I-T^\lambda(G)$ and $B^K$ coincide. This concludes the proof of Proposition~\ref{prop:KWKF}.
\end{proof}

As mentioned in the remark at the end of Subsection~\ref{sub:Fis}, the correspondence $G\mapsto\G_G$ used here is a variation of Fisher's original construction~\cite{Fi2},
that could have been used just as well. Assuming that all the vertices of $G$ are of even degree, there is yet another variation of the Fisher correspondence where each
vertex of degree $n$ gives rise to $2n-2$ vertices (see~\cite{D96}). It is also possible to use this version of the correspondence: the proof of Proposition~\ref{prop:KWKF}
only requires minor adjustments to be adapted to such variations on the theme. This yields an identification of the Kac-Ward method with the particular version of the Pfaffian method described in~\cite{D96}. Such a result is by no mean obvious: the special case of the planar square lattice alone is quite non-trivial, being the subject of the
article~\cite{Zin}.

\bibliographystyle{amsplain}
\bibliography{KW}

\end{document}